\documentclass[twocolumn,           
               showpacs,            
               nopreprintnumbers,     
               aps,                 
               prd,          	    
               a4paper,             
               groupedaddress,  
               unsortedaddress,
               nofootinbib,         
               tightenlines,        
               floats               
               ]{revtex4}

\usepackage{graphicx}
\usepackage{amssymb,latexsym}
\usepackage[draft=false]{hyperref}

\newcommand{\ba}{\begin{eqnarray}}
\newcommand{\ea}{\end{eqnarray}}  
\newcommand{\be}{\begin{equation}}
\newcommand{\ee}{\end{equation}}
\newcommand{\nn}{\nonumber \\}

\begin{document}

\hyphenation{brane-world}  




\title{Inflationary slow-roll formalism and perturbations in the\\ 
Randall--Sundrum Type II braneworld}
\author{Erandy Ram\'{\i}rez and Andrew R.~Liddle}
\affiliation{Astronomy Centre, University of Sussex, 
             Brighton BN1 9QJ, United 
Kingdom}
\date{\today} 
\pacs{98.80.Cq \hfill astro-ph/0309608}
\preprint{astro-ph/0309608}


\begin{abstract}
We formalize the Hubble slow-roll formalism for inflationary dynamics in 
Randall--Sundrum Type II braneworld cosmologies, defining Hubble slow-roll 
parameters which can be used along with the Hamilton--Jacobi formalism. 
Focussing on the high-energy limit, we use these to calculate the exact power 
spectrum for power-law inflation, and then perturb around this solution to 
derive the higher-order expression for the density perturbations (sometimes 
called the Stewart--Lyth correction) of slow-roll braneworld models. Finally we 
apply our result to specific examples of potentials to calculate the correction 
to the amplitude of the power spectrum, and compare it with the standard 
cosmology. We find that the amplitude is not changed significantly by the 
higher-order correction.   
\end{abstract}

\maketitle

\section{Introduction}

New ideas in fundamental physics have suggested the possibility that our
observable Universe lies on a `brane' within a higher-dimensional bulk 
space-time,
and this idea may have serious ramifications for early Universe cosmology
\cite{revs}.  The most studied scenario, viewed as a toy model for more
elaborate proposals, is the Randall--Sundrum Type II (RS-II) model where there
is a single brane living in an anti de Sitter bulk \cite{RSII}.  In that case,
the Friedmann equation is modified at high energy, potentially with significant
consequences for our understanding of inflation and the perturbations it
generates.

The most powerful tool for studying inflationary dynamics is the slow-roll
approximation \cite{SR}, which is expected to be extremely accurate for models
capable of matching current observations such as those of WMAP \cite{wmap}.
This enables considerable analytical progress to be made, and predictions for
the perturbations generated by inflation are most conveniently expressed in
terms of slow-roll parameters which measure the accuracy of the approximation
\cite{LL}.  Such an approach was used by Maartens et al.~\cite{MWBH} to describe
the outcome of inflation in the RS-II braneworld scenario.

In that paper, the version of the slow-roll approximation used was based on
derivatives of the potential driving inflation.  In this paper we formulate the
approximation in terms of derivatives of the Hubble parameter, which in the
standard cosmology is an important tool for developing high-accuracy predictions
for density perturbations.  We take advantage of our formalism to carry out
similar calculations for the high-energy limit of RS-II inflation models, 
deriving an exact result for 
power-law inflation and calculating the higher-order
slow-roll correction to the perturbation amplitude (often called the
Stewart--Lyth correction, as they carried out the original calculation for the
standard cosmology \cite{SL}).  We find that the correction is of a similar size
to that in the standard cosmology (i.e.~negligible in many inflationary
models), and calculate it for some sample inflation models.

\section{Slow-roll formalism}

We follow the notation set down by Liddle and Taylor \cite{LT}. In the
Randall--Sundrum Type II model \cite{RSII} the Friedmann equation
receives an additional term quadratic in the density \cite{HE}.  The
Hubble parameter $H$ is related to the energy density $\rho$ by
\begin{equation}
\label{Hubble}
H^2 = \frac{8\pi}{3 M_4^2} \, \rho \, \left(1 + \frac{\rho}{2\lambda} \right) 
\,,
\end{equation}
where $M_4$ is the four-dimensional Planck mass 
and $\lambda$ is the brane tension. We have set the four-dimensional 
cosmological constant to zero, and assumed that inflation rapidly
makes any dark 
radiation term negligible.
This reduces to the usual Friedmann equation 
for \mbox{$\rho \ll \lambda$}. 
We assume that the 
scalar 
field is confined to the brane, so it obeys the usual equation
\begin{equation}
\ddot{\phi}+ 3H \dot{\phi} = -V' \,,
\end{equation}
where prime indicates derivative with respect to $\phi$, and dot a 
derivative with respect to time. Its energy density is $\rho = 
V+\dot{\phi}^2/2$.

For the standard cosmology, the Hubble slow-roll formalism was set down in 
detail in Ref.~\cite{LPB}. The first two parameters are defined as
\be
\epsilon_{\rm H}  \equiv  3 \frac{\dot{\phi}^2/2}{V+\dot{\phi}^2/2}  = 
\frac{M_4^2}{4\pi}\, \frac{H'^2}{H^2} \,; 
\label{eq:ep0}
\ee
\be
\eta_{\rm H}  \equiv  - 3 \frac{\ddot{\phi}}{3H\dot{\phi}}  =
\frac{M_4^2}{4\pi}\, \frac{H''}{H}  \,.
\label{eq:et0}
\ee
If these parameters are much less than one, they allow the neglect of the 
$\dot{\phi}$ term in the Friedmann equation, and the $\ddot{\phi}$
term in the 
scalar wave equation \cite{SR}. In addition, the condition for inflation, 
$\ddot{a} >0$ is conveniently expressed as $\epsilon_{{\rm H}}<1$.

We seek a generalization of those parameters suitable for use with the 
braneworld equations. We note that there is some arbitrariness to the 
definitions, at least in terms of the constant prefactor. However this can be 
removed by imposing the requirements that the parameters
$\epsilon_{{\rm H}}$ and $\eta_{{\rm H}}$  continue to correspond to 
the conditions enabling 
neglect of terms in the Friedmann and fluid equations respectively, that 
$\epsilon_{\rm H} < 1$ remains the condition for inflation, and that in the 
slow-roll 
limit the density perturbation spectral index takes the same form as in the 
standard cosmology.

We will make use of an approach developed by Hawkins and Lidsey \cite{hl1,hl}, 
who devised a formalism for braneworld inflation with many of the properties 
of the Hamilton--Jacobi approach used in the standard cosmology \cite{SB}. They 
define a quantity $y(\phi)$, 
which is to 
play a similar role to $H(\phi)$ in the standard cosmology, by 
\be
y^2={{\rho/{2\lambda}}\over{1+\rho/{2\lambda}}} \,.
\label{eq:ratio}
\ee 
The inverse relation is
\be
\rho\equiv {{2\lambda y^2}\over{1-y^2}} \,.
\label{eq:y}
\ee
At low energies $\rho\ll\lambda$,
$y^2\rightarrow 0$, while at high energies $\rho\gg\lambda$,
$y^2\rightarrow 1$.

The Friedmann equation can be written as
\be
H(y)={\left({{16\pi\lambda}\over{3M_4^2}}\right)}^{1/2}{{y}\over{1-y^2}}\,.
\label{eq:fe1}
\ee
From the scalar wave equation and Eq.~(\ref{eq:y}) one obtains
\be
\dot{\phi}=-{\left({{\lambda M_4^2}\over{3\pi
}}\right)}^{1/2}{{y'}\over{1-y^2}} \,.
\label{eq:phi1}
\ee
Taking the derivative with respect to the field in the Friedmann
equation, one finds
\be
H'={\left({{16\pi\lambda}\over{3M_4^2}}\right)^{1/2}}{\left[{{1+y^2}\over{(1-y^2
) 
^2}}\right]}y' \,,
\label{eq:hp1}
\ee
and using Eq.~(\ref{eq:phi1})
\be
H'=-{{4\pi}\over{M_4^2}}{{(1+y^2)}\over{(1-y^2)}}\dot{\phi} \,.
\label{eq:hp2}
\ee

In Ref.~\cite{hl}, Hawkins and Lidsey define two parameters 
\begin{equation}
\beta \equiv \frac{M_4^2}{4\pi} \, \frac{y'^2}{y^2} \quad ; \quad \gamma \equiv 
\frac{M_4^2}{4\pi} \, \frac{y''}{y^2} \,,
\end{equation}
by analogy to the standard cosmology Hubble slow-roll parameters. These 
parameters prove useful in analyzing the exact dynamics of braneworld 
inflationary models. However, their smallness (as compared to unity) does not 
precisely correspond to the ability to neglect terms in the Friedmann and scalar 
wave equations, and this means they are not ideal for the purpose of analyzing 
perturbation generation. We therefore define new Hubble slow-roll parameters for 
the braneworld, which do have this property.

To do this, we define the parameters as ratios of terms in the Friedmann and 
wave equations, following Eqs.~(\ref{eq:ep0}) and (\ref{eq:et0}):
\be   
\epsilon_{\rm H}=C(y){{\dot{\phi^2}/2}\over{V+\dot{\phi}^2/2}}\quad ;\quad
\eta_{\rm H}=-D(y){{\ddot{\phi}}\over{3H\dot{\phi}}} \,,
\label{eq:srp2}
\ee
At this stage we have allowed the `constant' prefactors $C(y)$ and $D(y)$ to 
depend on $y$; we will next show how to fix them using the requirements that 
$\epsilon_{{\rm H}}<1$ corresponds to inflation, and that in the slow-roll limit 
the density perturbation spectral index takes its usual form. In order for these 
parameters to correspond to the ability to neglect terms, those prefactors 
should always be of order unity, and we will soon see that they are.

From the definition of $\epsilon_{\rm H}$, using Eqs.~(\ref{eq:fe1}) and 
(\ref{eq:hp2}), one can find
\be
\epsilon_{\rm H}={{C(y)}\over{3}}{\left({{\lambda 
M_4^2}\over{3\pi}}\right)^{1/2}}{{y}\over{(1+y^2)^2}}{{H'^{2}}\over{H^3}}.
\label{eq:e2}
\ee
The coefficient $C(y)$ can be determined by demanding that
$\ddot{a}=0 \Longleftrightarrow \epsilon_{{\rm H}}=1$; taking the derivative of 
the Friedmann equation with 
respect to time gives
\be
\frac{\ddot{a}}{a}=\frac{16\pi\lambda
y^2}{M_4^2(1-y^2)^2}\left[\frac{-(1+y^2)}{C(y)}+\frac{1}{3}\right]
\label{eq:acc}
\ee
and so we require $C(y) \equiv 3(1+y^2)$. Its value ranges from 3 in the 
low-energy limit to 6 in the high-energy limit. Our definition for 
$\epsilon_{\rm H}$ is therefore
\be
\epsilon_{\rm H} \equiv {\left({{\lambda
M_4^2}\over{3\pi}}\right)^{1/2}}{{y}\over{(1+y^2)}}{{H'^{2}}\over{H^3}} \,.
\label{eq:ep4}
\ee
In the low-energy limit, Eq. (\ref{eq:ep4}) becomes
the usual expression Eq.(\ref{eq:ep0}). 

For $\eta_{\rm H}$ we apply Eqs.~(\ref{eq:fe1}) and (\ref{eq:hp2}) to obtain
\be
\eta_{\rm H}={{D(y)}\over{3}}\left({{\lambda M_4^2}\over{3\pi}}\right)^{1/2}
\left[{{{y}\over{(1+y^2)}} \frac{H''}{H^2}-{{4y^3}\over{(1+y^2)^3}}} 
\frac{H'^2}{H^3} \right],
\label{eq:et2}
\ee
There are several ways one could aim to fix the constant $D(y)$ to establish a 
unique definition of $\eta_{{\rm H}}$. We choose to do so such that the 
slow-roll expression for the density perturbation spectral index takes its usual 
form, namely $n=1-4\epsilon_{\rm H}+2\eta_{\rm H}$, as was done for the 
potential slow-roll parameters in Ref.~\cite{MWBH}. The slow-roll expression for 
the perturbation amplitude is \cite{MWBH}
\begin{equation}
\label{e:sramp}
{\cal P}_{\cal R}^{1/2} = \left. \frac{H^2}{2\pi \dot{\phi}} \right|_{k=aH} \,,
\end{equation}
which in terms of those variables is exactly the usual result. The spectral 
index is defined as
\begin{equation}
n-1 = \frac{d \ln {\cal P}_{\cal R}}{d\ln k} \,,
\end{equation}
and using Eqs.~(\ref{eq:fe1}) and (\ref{eq:hp2}) one obtains
\ba
n-1 & =& -4{\left({{\lambda
M_4^2}\over{3\pi}}\right)^{1/2}}{{y}\over{(1+y^2)}}{{H'^{2}}
\over{H^3}}+ \\
&&\hspace*{-12pt} +2\left({{\lambda M_4^2}\over{3\pi}}\right)^{1/2}\left[{{{y}
\over{(1+y^2)}}\, \frac{H''}{H^2} -{{4y^3}\over{(1+y^2)^3}}} \, 
\frac{H'^2}{H^3}\right]\,, \nonumber
\label{eq:si}
\ea
This agrees with the usual expression $n=1-4\epsilon_{\rm H}+2\eta_{\rm H}$ 
provided $D(y)$ is set to $3$ for
all regimes. That gives us our definition
\be
\eta_{\rm H} \equiv\left({{\lambda M_4^2}\over{3\pi}}\right)^{1/2}\left[{{{y}
\over{(1+y^2)}}\, \frac{H''}{H^2} - {{4y^3}\over{(1+y^2)^3}}} \, 
\frac{H'^2}{H^3} \right] \,.
\label{eq:et3}
\ee
This too reduces to the usual expression, Eq.~(\ref{eq:et0}), in the low-energy 
regime. 

The two definitions Eqs.~(\ref{eq:ep4}) and (\ref{eq:et3}) define Hubble 
slow-roll parameters valid in all regimes of RS-II brane inflation, generalizing 
the usual ones while preserving many key results: they give the conditions for 
neglecting terms in the Friedmann and fluid equations, $\epsilon_{{\rm H}} < 1$ 
corresponds to an inflationary expansion, and the slow-roll spectral index 
formula is always $n=1-4\epsilon_{\rm H}+2\eta_{\rm H}$.

\section{Exact and higher-order perturbations in the high-energy regime}

In this section we exploit the formalism of the previous section to make 
accurate calculations of the density perturbations. Throughout this section we 
will restrict ourselves to the high-energy regime, obtained by taking $y 
\rightarrow 1$, where our slow-roll parameters can be written
\begin{equation}
\label{eq:hesr}
\epsilon_{{\rm H}} = \frac{M_5^3}{4\pi} \, \frac{H'^2}{H^3} \quad ; \quad
\eta_{{\rm H}} + \epsilon_{{\rm H}} =\frac{M_5^3}{4\pi} \, \frac{H''}{H^2}
\end{equation}
where $M_5 \equiv (4\pi \lambda/3)^{1/6} M_4^{1/3}$ is the five-dimensional 
Planck mass. The high-energy versions of Eqs.~(\ref{eq:hp2}) and (\ref{eq:fe1}) 
are
\be
\dot{\phi}=-\frac{M_5^3}{4\pi}\frac{H'}{H}\,,
\label{eq:hephif}
\ee
and 
\be
H=\frac{4\pi}{3M_5^3}\rho \,.
\label{eq:hefe}
\ee
All these expressions could have been obtained directly for the high-energy 
regime using the same criteria we set down for the general case in the previous 
section.

We will consider one case where the perturbations can be obtained exactly 
(namely power-law inflation, though in this case the corresponding potential is 
not exponential), and then carry out our main calculation which is to compute 
the
correction to the density perturbation amplitude from next-order in slow-roll.
This type of calculation was first performed by Stewart and Lyth for the
standard cosmology \cite{SL}, and is often known as the Stewart--Lyth
correction.  We will compute its equivalent for the density
perturbations 
in the
high-energy regime of the RS-II model.

We will calculate the perturbations using a formalism due to 
Mukhanov \cite{muk}. He defined a new variable $u=a\delta\phi$ (where 
$\delta\phi$ is defined in the spatially-flat gauge; a gauge-invariant 
definition can be made which includes a contribution from the curvature 
perturbation), and demonstrated 
that in linear perturbation theory its Fourier modes obey the wave equation
\begin{equation}
\label{e:modeeq}
\frac{d^2u_k}{d\tau^2} + \left( k^2 - \frac{1}{z}\, \frac{d^2z}{d\tau^2} \right) 
\, u_k = 0 \,.
\end{equation}
Here $\tau$ is the conformal time, and $z \equiv a\dot{\phi}/H$
encodes all the relevant information about how the background is evolving. This 
equation has been fully derived only in the standard cosmology, and we 
stress that it is currently a conjecture that it can still be 
used in the braneworld context. This is because as yet no-one has been able to 
calculate 
the backreaction from five-dimensional gravity in order to assess whether it is 
different from the four-dimensional backreaction that the Mukhanov equation 
incorporates, other than in the large-scale limit where energy conservation 
alone is sufficient to ensure the perturbations remain constant \cite{WMLL}. 
There is however supporting evidence for its use, because its derivation 
does not need the Friedmann equation. This is clear since the Mukhanov equation 
does not feature the gravitational constant $G$ (though some derivations of it 
do appear to use the Friedmann equation in intermediate steps), and we have 
verified explicitly that the same equation does result using the modified 
Friedmann equation in the derivation as given in Ref.~\cite{LLbook}. In other 
words, 
the Mukhanov equation tells us how scalar field perturbations develop in a given 
expanding background, without needing to know the physics responsible for 
determining the background evolution. This however falls short of being a full 
five-dimensional calculation as would be required to fully verify the use of the 
equation, but as yet it is not known how to implement such a calculation.

Having adopted the Mukhanov equation, the first step is to write it in terms of 
the slow-roll parameters, with a lengthy calculation yielding
\ba
\frac{1}{z}\, \frac{d^2z}{d\tau^2} =
2a^2H^2\left[1+\epsilon_{\rm H}-{{3}\over{2}}\eta_{\rm
H}-{{7}\over{2}}\epsilon_{\rm H}\eta_{\rm H}+ \right. \nn
\left. +{{1}\over{2}}\eta_{\rm H}^2
+{{M_5^6}\over{32\pi^2}}{{H'H'''}\over{H^4}}\right] \,,
\label{eq:ex}
\ea
which is an exact relation.

\subsection{Exact mode equation solution}

It is well known that the mode function can be solved exactly if the 
square-bracketted term in Eq.~(\ref{eq:ex}) is constant, which in the standard 
cosmology corresponds to power-law inflation from an exponential potential 
\cite{LM}, the 
calculation having first been performed by Lyth and Stewart \cite{LS}. To 
discover if there is an analogous result for the RS-II scenario, we set 
$\epsilon_{{\rm H}} = 1/p$ where $p>1$ is a constant, and examine whether this 
makes the square bracket constant.

Taking the derivative of $\epsilon_{\rm H}$ with respect to the field,
one can write 
\be
\epsilon_{{\rm H}}'= \frac{H'}{H}\left[2\eta_{{\rm H}}-\epsilon_{{\rm H}}
\right] \,.
\label{eq:etaepsilonp}
\ee
As we have demanded $\epsilon_{{\rm H}}$ is constant, this implies 
\mbox{$\eta_{{\rm H}} = 1/2p$}. Similarly, differentiating $\eta_{{\rm H}}$ 
gives  
\ba
\frac{M_5^6}{32\pi^2}\frac{H'H'''}{H^4}=\epsilon_{\rm H}\left[(\eta_{\rm
H}+\epsilon_{\rm H})+\frac{H}{2H'}(\eta_{\rm H}'+\epsilon_{\rm H}')\right]
\ea
implying
\be
\frac{M_5^6H'H'''}{32\pi^2H^4}=\frac{3}{2p^2}\,.
\label{eq:lp}
\ee
The square-bracket of Eq.~(\ref{eq:ex}) therefore indeed is constant, so the 
equation can be solved exactly.

Before going on to do that, it is interesting to ask what potential gives this 
solution. Solving for the Hubble
parameter from the definition of $\epsilon_{{\rm H}}$, and then substituting 
into the Hamilton--Jacobi equation in the
high-energy limit, namely
\be
H-{{{M_5^3}\over{24\pi}}{{H^{'2}}\over{H^2}}}={{4\pi}\over{3M_5^3}}V(\phi)\,,
\label{eq:hjhe}
\ee
we find that the corresponding potential is 
\be
V(\phi)=\frac{1}{8}\frac{M_5^6(6p-1)}{\pi^2\phi^2}\,.
\label{eq:pot}
\ee 
Instead of the exponential potential found in the standard cosmology, we have an 
inverse power-law potential. Nevertheless, the corresponding expansion law $a 
\propto t^p$ is power-law inflation as usual.

Following Refs.~\cite{LS,SL,rec}, the conformal time for constant
$\epsilon_{\rm H}$ is given by
\be
\tau=-{{1}\over{aH}}{{1}\over{1-\epsilon_{\rm H}}},
\label{eq:tauex}
\ee
and Eq.~(\ref{eq:ex}) with the values of the Hubble slow-roll
parameters gives 
\begin{equation}
\frac{1}{z}\, \frac{d^2z}{d\tau^2}= \frac{1}{4}\frac{8p^2+2p-1}{\tau^2
(p-1)^2} \,.
\end{equation}
This allows one to write the Mukhanov equation Eq.~(\ref{e:modeeq}) 
as a Bessel-like one
\be
\left[\frac{d^2}{d\tau^2}+k^2-\frac{(\nu^2-1/4)}{\tau^2}\right]u_k=0,
\ee
with $\nu=3p/2(p-1)$. The solution with the appropriate behaviour at small 
scales can be written as
\be
u_k(\tau)=\frac{\sqrt{\pi}}{2}e^{i(\nu+1/2)\pi/2}(-\tau)^{1/2}
H_{\nu}^{(1)}(-k\tau), \label{eq:s1}
\ee
where $H_{\nu}^{(1)}$ is the Hankel function of the first kind of
order $\nu$. The asymptotic form of this equation once the mode is
outside of the horizon is obtained by taking the limit
$k/aH\rightarrow 0$
\be
u_k\rightarrow
e^{i(\nu-1/2)\pi/2}2^{\nu-3/2}\frac{\Gamma(\nu)}{\Gamma(3/2)}\frac{1}{\sqrt{2k}}
(-k\tau)^{-\nu+1/2},
\label{eq:s2}
\ee
from which the corresponding form of the power spectrum using \cite{rec}
\begin{equation}
{\cal P}_{\cal
R}^{1/2}(k)=\sqrt{\frac{k^3}{2\pi^2}}\left|\frac{u}{z}\right|\,
\end{equation}
yields
\be
{\cal P}_{\cal 
R}^{1/2}(k)=\frac{3^{\nu-1/2}\nu^{1/2-\nu}}{M_5^3}\frac{\Gamma(\nu)}{\Gamma(3/2)
} \left. \frac{H^3}{|H'|}\right|_{k=aH}
\label{eq:ps1}
\ee

\subsection{Higher-order perturbation calculation}

We now perturb around the exact solution given above for small $\epsilon_{{\rm 
H}}$ and $\eta_{{\rm H}}$, following 
Refs.~\cite{SL} and \cite{rec}. The expansion to lowest order of the
conformal time gives 
\be
\tau=-\frac{1}{aH}(1+\epsilon_{\rm H}).
\label{eq:ct2}
\ee
Applying this in Eq.~(\ref{eq:ex}) and truncating the expansion to
first order, one arrives at another Bessel equation, now with $\nu$ given by 
\begin{equation}
\nu\simeq \frac{3}{2}+2\epsilon_{\rm H}-\eta_{\rm H}\,.
\end{equation}
Note that the final three terms of Eq.~(\ref{eq:ex}) do not affect the form of 
this expression. Then the solution
Eq.~(\ref{eq:s2}) can be used with the new form for $\nu$ and the conformal
time, expanding also the Gamma function and the other expressions to
first-order, to obtain the final result
\begin{equation}
{\cal P}_{{\cal R}}^{1/2} = \left[ 1 - (2C+1) \epsilon_{\rm H} + C
\eta_{\rm H} \right] \, 
\left. \frac{2H^3}{M_5^3 |H'|} \; \right|_{k=aH} 
\label{eq:ps}
\end{equation}
where $C = -2+\ln 2 + b \simeq -0.73$, with $b$ the Euler-Mascheroni 
constant. 
The leading-order term, obtained by setting the square bracket to one, agrees 
with the result of Maartens et al.~\cite{MWBH} in the high-energy limit.

While formally the correction term in the square-bracket looks exactly the same
as in the standard cosmology \cite{SL,rec}, we should recall that the slow-roll
parameters which appear in it are generalizations of those in the standard
cosmology, and could have been defined in different ways.  To get a feel for
what this correction means, we need to evaluate it for some characteristic
potentials, which we do in the next subsection.

\subsection{Specific examples}

To determine the typical size of the  next-order correction,
we study the monomial potentials $V \propto \phi^\alpha$ 
for $\alpha = 2$, 
$4$ and $6$, assuming inflation takes place well within the high-energy 
regime. For comparison, we also calculate the magnitude of the correction for 
the standard cosmology.

To calculate the size of the correction term, we can use the slow-roll 
approximation for $\epsilon_{{\rm H}}$ and $\eta_{{\rm H}}$, since any 
corrections to that will be of higher-order. The simplest approach is to 
rewrite $\epsilon_{\rm H}$ and  
$\eta_{\rm H}$ in terms of the potential and its derivatives. We will make use 
of
the equations
\begin{eqnarray}
H & \simeq & {{4\pi}\over{3M_5^3}} \, V \,; \label{eq:fsphieq1}\\
3H\dot{\phi} & \simeq & -V^{\prime} \,;\\
\dot{\phi} & = & -\frac{M_5^3}{4\pi}{{H^{\prime}}\over{H}}\,.
\label{eq:fsphieq}
\end{eqnarray}
The first and second of these use the slow-roll approximation, and the first and 
third use the high-energy approximation. These enable us to obtain the relations 
\begin{equation}
\epsilon_{\rm H} \simeq \epsilon_{{\rm V}} \quad ; \quad
\eta_{\rm H} \simeq \eta_{{\rm V}} - \epsilon_{{\rm V}}\,,
\end{equation}
where the potential slow-roll parameters are
\begin{equation}
\epsilon_{\rm V}={{{3M_5^6}\over{16\pi^2}}{V'^2\over{V^3}}}\, \quad ; \quad
\eta_{\rm V}={{{3M_5^6}\over{16\pi^2}}{{V''}\over{V^2}}}\,,
\label{eq:srpv}
\end{equation}
being the high-energy limit of the parameters as defined by Maartens et 
al.~\cite{MWBH}. Using them, we can write Eq.~(\ref{eq:ps}) as
\be
{\cal P}_{\cal R}^{1/2}={[1-(3C+1)\epsilon_{\rm V}+C\eta_{\rm 
V}]} \left. {{2H^3}\over{M_5^3|H'|}}\right|_{k=aH} \,,
\label{eq:psv}
\ee 
where we aim to calculate the square-bracketted term.

We assume that observable scales crossed the Hubble radius 50 $e$-foldings 
before the end of inflation, and need to compute the slow-roll parameters at 
that time. We take the potential as
\be
V(\phi) = m \phi^{\alpha}\,,
\label{eq:p}
\ee
where $m$ is a constant. Putting this in Eq.~(\ref{eq:srpv}) 
and 
setting  $\epsilon_{{\rm V}}=1$, which corresponds to the end of inflation, we 
can obtain
the value of $\phi_{\rm end}$. We use this in the expression for the number
of $e$-foldings in the high-energy limit \cite{MWBH}
\begin{equation}
N\simeq -{{16\pi^2}\over{3M_5^6}}\int_{\phi_N}^{\phi_{{\rm 
end}}}{{V^2}\over{V'}}d\phi\,.
\end{equation}
Taking $N=50$, this gives the value of $\phi_{50}$ and substituting
in the equations for $\epsilon_{\rm V}$ and $\eta_{\rm V}$ one gets 
\ba
\epsilon_{{\rm V},50} & = & {{\alpha}\over{100+51\alpha}} \,;\\
\eta_{{\rm V},50} & =& {{\alpha-1}\over{100+51\alpha}} \,.
\label{eq:srp50}
\ea

In the case of standard cosmology, the same calculation is carried out using the 
corresponding expressions for the slow-roll parameters
and the power spectrum given by Refs.~\cite{SL,rec} 
\be
{\cal P}_{\cal
R}^{1/2}(k)=[1-(2C+1)\epsilon_{\rm H}+C\eta_{\rm 
H}]\left.\frac{2H^2}{M_4^2|H'|}\right|_{k=aH}
\label{eq:pssc}
\ee
with $\epsilon_{\rm H}$ and $\eta_{\rm H}$, defined as in Eqs.~(\ref{eq:ep0}) 
and 
(\ref{eq:et0}), being the  Hubble slow-roll parameters in the standard 
cosmology.
To rewrite them in terms of the potential, we use the equations \cite{rec}
\ba
H^2 & \simeq & \frac{8\pi}{3M_4^2} \,V \,;\\
3H\dot{\phi} & \simeq & -V'\,;\\
\dot{\phi}  & = & - {\frac{M_4^2}{4\pi}H'} \,,
\label{eq:fsphisc}
\ea
which are the standard cosmology equivalents of 
Eqs.~(\ref{eq:fsphieq1})--(\ref{eq:fsphieq}). This leads to the same relations 
as in
the previous case
\begin{equation}
\epsilon_{\rm H} \simeq \epsilon_{{\rm V}} \quad ; \quad
\eta_{\rm H} \simeq \eta_{{\rm V}} - \epsilon_{{\rm V}}\,,
\end{equation}
with
\be
\epsilon_{{\rm V}}=\frac{M_4^2}{16\pi} \,\frac{V'^2}{V^2} \quad ;
\quad \eta_{{\rm V}}=\frac{M_4^2}{8\pi}\frac{V''}{V},
\label{eq:ssrp}
\ee
Then the power spectrum can be written as
\be
{\cal P}_{\cal
R}^{1/2}(k)=[1-(3C+1)\epsilon_{\rm V}+C\eta_{\rm 
V}]\left.\frac{2H^2}{M_4^2|H'|}\right|_{k=aH}
\ee
where now $N\simeq -(8\pi/M_4^2)\int_{\phi_N}^{\phi_{{\rm end}}}(V/V')d\phi$.

\begingroup
\begin{table}[t]
\begin{center}
\renewcommand*{\arraystretch}{2}
\caption{Fractional correction to the power spectrum.}
\vspace*{6pt}
\begin{tabular}{l|ccc|ccc|}
& \multicolumn{3}{c|}{standard cosmology} & \multicolumn{3}{c|}{high-energy 
braneworld} \\
& $\epsilon_{{\rm V},50}$
& $\eta_{{\rm V},50}$
& correction
& $\epsilon_{{\rm V},50}$
& $\eta_{{\rm V},50}$
& correction
\\
\colrule
$\alpha=2$
&$\frac{1}{101}$
&$\frac{1}{101}$
&$1.0046$
&$\frac{1}{101}$
&$\frac{1}{202}$
&$1.0082$
\\
$\alpha=4$
&$\frac{1}{51}$
&$\frac{3}{102}$
&$1.0019$
&$\frac{1}{76}$
&$\frac{3}{304}$
&$1.0085$
\\
$\alpha=6$
&$\frac{3}{103}$
&$\frac{5}{103}$
&$0.9992$
&$\frac{3}{203}$
&$\frac{5}{406}$
&$1.0086$
\label{table:bw}
\end{tabular}

\end{center} 
\end{table}
\endgroup

The results for both regimes are shown in Table~\ref{table:bw}.
They show that the magnitude of the correction is similar in both cases, 
though it differs in detail. 
Nevertheless, these 
results confirm that the amplitude of the power spectrum is not
changed significantly with respect to the slow-roll result by the higher-order 
correction.

\subsection{Correction from the high-energy approximation}

We end by mentioning that, especially when it is small, the higher-order 
slow-roll 
correction may well be sub-dominant to corrections coming from the high-energy 
approximation not being exact. The slow-roll approximation allows us to 
determine the size of such corrections; using Eq.~(\ref{e:sramp}) which is valid 
in any regime, we find the slow-roll perturbation amplitude to be 
\begin{equation}
{\cal P}_{{\cal R}}^{1/2} \simeq
\left. \left[  1+\frac{1}{2} \, \left(\frac{\lambda}{V}\right)^2 \right] 
\,\frac{2H^3}{M_5^3 |H'|} \; \right|_{k=aH}  \,,
\label{e:heapp}
\end{equation}
where we have expanded the result in terms of the parameter $\lambda/V$ which is 
small in the high-energy regime. Because the term linear in $\lambda/V$ happens 
to cancel, the high-energy approximation is a better one than might have been 
expected, and we only need $V \gtrsim 10\lambda$ to bring the correction from 
that approximation within one percent. More generally, comparison of 
Eqs.~(\ref{eq:ps}) and (\ref{e:heapp}) allows a test of when the higher-order 
slow-roll correction dominates the correction to the high-energy approximation.

If one were very ambitious, one could also attempt to generalize our 
higher-order 
slow-roll calculation to be valid in any regime, but given the limited ability 
of observations to probe or distinguish amongst such small corrections, such a 
calculation does not seem worthwhile. The same remark applies to an attempt to 
calculate the higher-order gravitational wave spectrum correction; as 
gravitational waves are known to be subdominant such corrections are even less 
relevant, and also much harder to calculate due to the gravitational waves' 
ability to penetrate the bulk dimension.

\section{Conclusions}

We have devised a Hubble slow-roll formalism for inflation in the RS-II 
braneworld cosmology, extending work by Hawkins and Lidsey \cite{hl1,hl} to 
define parameters which share the nice properties of those used in the 
standard cosmology, which are recovered in that limit. As an application, we 
have computed the density perturbation spectrum in the high-energy limit, both 
exactly for power-law inflation and to higher-order for general slow-roll 
inflation models. To do so we have used the Mukhanov equation; while no one has 
yet been able to prove that this equation is still valid in the braneworld 
context, we have provided some evidence supporting its use.
We have also quantified how well the high-energy approximation 
must hold in order for the higher-order slow-roll correction to be the dominant 
one.

It is interesting to note that, having defined the slow-roll parameters 
$\epsilon_{{\rm H}}$ and $\eta_{{\rm H}}$ so as to give the usual spectral index 
formula for slow-roll perturbations, it turns out that the next-order correction 
is of the same form as in the standard cosmology. We are not aware of a 
physical reason which leads to this result. Nevertheless, for a given choice of 
potential one expects that observable perturbations are generated at a different 
location on that potential depending on the braneworld regime, and so 
predictions for both the spectral indices (see e.g.~Ref.~\cite{LidS}) and for 
the higher-order corrections will be different. We have examined the magnitude 
of the correction for some simple potentials, and we conclude that there is no 
reason to believe that the higher-order correction might be more important in 
the high-energy regime than in the standard cosmology. As recent observations 
including WMAP have restricted 
viable inflation models to regions close to the slow-roll limit, such 
corrections are expected to be small.

\begin{acknowledgments}
E.R. was supported by Conacyt and A.R.L. in part by the Leverhulme Trust.
We thank James Lidsey, David Lyth, Karim Malik and David Wands for useful 
discussions. 
\end{acknowledgments}

 

\begin{thebibliography}{50}
\bibitem{revs} For reviews see F. Quevedo, Class. Quant. Grav. {\bf 19}, 
	5721 (2002), {\tt hep-th/0210292}; P. Brax and C. van de Bruck, 
	Class. Quant. Grav. {\bf 20}, R201 (2003), {\tt hep-th/0303095}.
\bibitem{RSII} L. Randall and R. Sundrum, Phys. Rev. Lett. {\bf 83}, 4690 
	(1999), {\tt hep-th/9906064}.
\bibitem{SR} P. J. Steinhardt and M. S. Turner, Phys. Rev. D {\bf 29}, 2162
	(1984); E. W. Kolb and M. S. Turner, {\em The Early Universe}, 
	Addison--Wesley, Redwood City (1990).
\bibitem{wmap} C. L. Bennett {\it et al.}, Astrophys. J. Supp. {\bf 148}, 1
	(2003), {\tt astro-ph/0302207}; D. N. Spergel {\it et al.}, 
	Astrophys. J. Supp. {\bf 148}, 175 (2003), {\tt astro-ph/0302209}; 
	H. V. Peiris {\it et al.}, Astrophys. J. Supp. {\bf 148}, 213 (2003),
	{\tt astro-ph/0302225}.
\bibitem{LL} A. R. Liddle and D. H. Lyth, Phys. Lett. B {\bf 291}, 391 (1992),
	{\tt astro-ph/9208007}.
\bibitem{MWBH} R. Maartens, D. Wands, B. A. Bassett, and I. P. C. Heard,
	Phys. Rev. D {\bf 62}, 041301 (2000), {\tt hep-ph/9912464}.
\bibitem{SL} E. D. Stewart and D. H. Lyth, Phys. Lett. B {\bf 302}, 171(1993),
	{\tt gr-qc/9302019}.
\bibitem{LT} A. R. Liddle and A. N. Taylor, Phys. Rev. D {\bf 65}, 041301
	(2002), {\tt astro-ph/0109412}.
\bibitem{HE} C. Cs\'aki, M. Graesser, C. Kolda, and J. Terning, Phys. Lett. 
	B{\bf 462}, 34 (1999), {\tt hep-ph/9906513}; J. M. Cline, C. Grojean, 
	and G. Servant, Phys. Rev. Lett. {\bf 83}, 4245 (1999), {\tt
	hep-ph/9906523}; P. Bin\'etruy,
	C. Deffayet, U. Ellwanger, and D. Langlois, Phys. Lett. B {\bf 477}, 
	285 (2000), {\tt hep-th/9910219}; T. Shiromizu, K. I. Maeda, and 
	M. Sasaki, Phys. Rev. D{\bf 62}, 024012 (2000), {\tt gr-qc/9910076}. 
\bibitem{LPB} A. R. Liddle, P. Parsons, and J. D. Barrow, Phys. Rev.
        D {\bf 50}, 7222 (1994), {\tt astro-ph/9408015}.
\bibitem{hl1} R. M. Hawkins and J. E. Lidsey, Phys. Rev. D{\bf 63},
        041301 (2001), {\tt gr-qc/0011060}.
\bibitem{hl} R. M. Hawkins and J. E. Lidsey, astro-ph/0306311.
\bibitem{SB} D. N. Salopek and J. R. Bond, Phys. Rev D{\bf 42}, 3936 (1990).
\bibitem{muk} V. F. Mukhanov, Pis'ma Zh. Eksp. Teor. Fiz. {\bf 41}, 402 
	(1985) [Sov. Phys. JETP Lett. {\bf 41}, 493 (1985)];
	V. F. Mukhanov, Phys. Lett. B {\bf 218}, 17 (1989); see also
	M. Sasaki, Prog. Theor. Phys. {\bf 76}, 1036 (1986).
\bibitem{WMLL} D. Wands, K. A. Malik, D. H. Lyth, and A. R. Liddle,
	Phys. Rev. D{\bf 62}, 043527 (2000), {\tt astro-ph/0003278}. 
\bibitem{LLbook} A. R. Liddle and D. H. Lyth, {\em Cosmological inflation and 
	large-scale structure}, Cambridge University Press, Cambridge (2000).
\bibitem{LM} F. Lucchin and S. Matarrese, Phys. Rev. D{\bf 32}, 1316 (1985).
\bibitem{LS} D. H. Lyth and E. D. Stewart, Phys. Lett. B{\bf 274}, 168 (1992).
\bibitem{rec} J. E. Lidsey, A. R. Liddle, E. W. Kolb, E. J. Copeland,
        T. Barreiro, and M. Abney, Rev.  Mod. Phys. {\bf 69},
        373, (1997), {\tt astro-ph/9508078}.
\bibitem{LidS} A. R. Liddle and A. J. Smith, Phys. Rev. D{\bf 68}, 061301(R)
	(2003), {\tt astro-ph/0307017}.
\end{thebibliography}
\end{document}